\documentclass{article}

\usepackage{microtype}
\usepackage{graphicx}
\usepackage{subfigure}
\usepackage{booktabs} 
\usepackage{color}
\usepackage{algorithm}
\usepackage{algorithmic}
\usepackage[algo2e]{algorithm2e}
\usepackage{hyperref}

\newcommand{\model}{\sc ADAPT}
\usepackage[accepted]{icml2024}

\usepackage{amsmath}
\usepackage{amssymb}
\usepackage{mathtools}
\usepackage{amsthm}
\usepackage{booktabs}
\usepackage{amsfonts,amssymb}
\usepackage{multirow}
\usepackage{graphicx}
\usepackage{verbatim}
\usepackage[capitalize,noabbrev]{cleveref}
\usepackage{enumitem}

\theoremstyle{plain}

\theoremstyle{definition}

\theoremstyle{remark}

\usepackage[textsize=tiny]{todonotes}
\setlist[itemize]{noitemsep, topsep=0pt, partopsep=0pt, parsep=0pt, itemsep=0pt, leftmargin=*}

\icmltitlerunning{ADAPT: Alzheimer's Diagnosis through Adaptive Profiling Transformers}

\begin{document}

\twocolumn[
\icmltitle{
ADAPT: Alzheimer's Diagnosis through Adaptive Profiling Transformers
}



\icmlsetsymbol{equal}{*}

\begin{icmlauthorlist}
\icmlauthor{Yifeng Wang}{uiuc}
\icmlauthor{Ke Chen}{uiuc}
\icmlauthor{Haohan Wang}{uiuc}
\end{icmlauthorlist}

\icmlaffiliation{uiuc}{University of Illinois at Urbana-Champaign}

\icmlcorrespondingauthor{Haohan Wang}{haohanw@illinois.edu}

\icmlkeywords{Machine Learning, ICML}

\vskip 0.3in
]



\printAffiliationsAndNotice{}  

\begin{abstract}
\label{sec:abstract}

Automated diagnosis of Alzheimer's Disease (AD) from brain imaging, such as magnetic resonance imaging (MRI),
has become increasingly important and has attracted the community to contribute many deep learning methods. 
However, many of these methods are facing a trade-off that 3D models tend to be complicated while 2D models
cannot capture the full 3D intricacies from the data.
In this paper, we introduce a new model structure for diagnosing AD, and it 
can complete with 3D model's performances while 
essentially is a 2D method (thus computationally efficient). 
While the core idea lies in new perspective of cutting the 3D images into multiple 2D slices from three dimensions, 
we introduce multiple components that can further benefit the model in this new perspective, 
including adaptively selecting the number of sclices in each dimension, and the new attention mechanism.
In addition, we also introduce a morphology augmentation, 
which also barely introduces new computational loads,
but can help improve the diagnosis performances 
due to its alignment to the pathology of AD. 
We name our method {\model}, which stands for Alzheimer’s Diagnosis through Adaptive Profiling Transformers.
We test our model from a practical perspective (the testing domains do not appear in the training one):
the diagnosis accuracy favors our {\model}, 
while {\model} uses less parameters than most 3D models use. 

\end{abstract}    
\section{Introduction}
\label{sec:intro}

\begin{figure}
\centering
\includegraphics[scale=0.36]{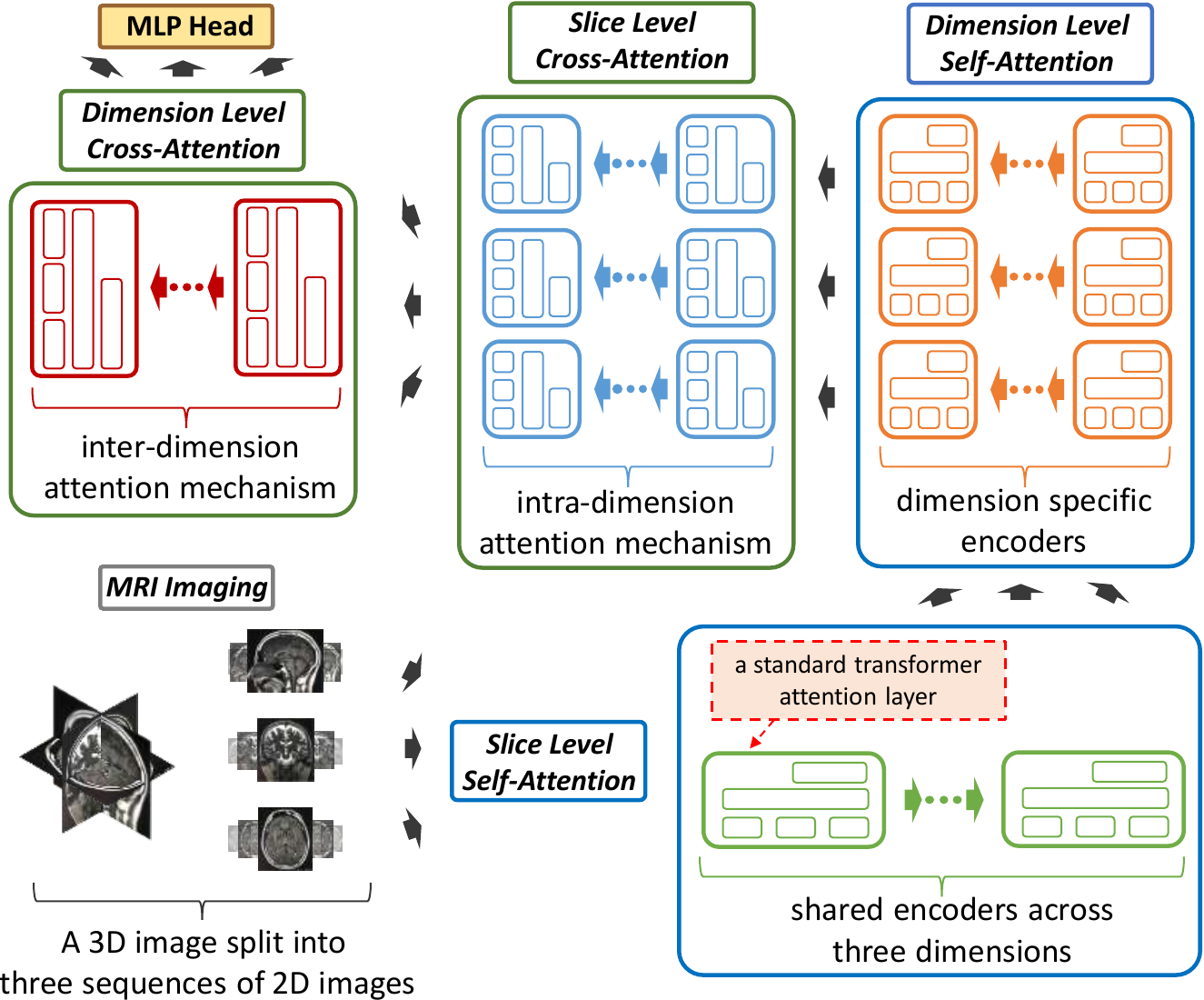}
  \caption{The overall framework of our proposed 2D ViT-Based Model for 3D Medical Imaging ({\model}).}
  \label{fig:teaser}
\end{figure}

Alzheimer's disease (AD) is a highly common neurodegenerative disorder that is usually diagnosed by structural alterations of the brain mass. 
Assessing an AD usually involves the acquisition of magnetic resonance imaging (MRI) images, since it offers accurate visualization of the anatomy and pathology of the brain~\cite{zhou2023learning}.
To overcome the vulnerability of misdiagnosis~\cite{despotovic2015mri} and to speed the diagnosis process,
the community has been using machine intelligence to help physicians diagnose AD diseases~\cite{jo2019deep}. 

Considering the complex structure of brain magnetic resonance imaging (MRI), in recent years, Convolutional Neural Networks (CNNs) have been established with a dominant performance in the AD-related field~\cite{salehi2020cnn,farooq2017deep}, due to their effectiveness in extracting meaningful spatial hierarchical features from complex images. 
Many  methods~\cite{zhu2021dual,wen2020convolutional} try to learn the characteristics of AD using CNN-based models. However, the original MRI is complex 3D data, with the proposed 3D model, the input of the 3D convolution operation introduces a third dimension, which greatly increases the burden on the computer.
So they use a bag of patches selected from the skull-stripped brain region. These approaches disregard the global context information, which can have a substantial impact on accurately identifying lesions during inference~\cite{wang2022super}. 
Moreover, CNNs are not well-suited for mining global long-dependent information due to their inherent focus on extracting local information~\cite{luo2016understanding,dosovitskiy2020image}.

Transformers~\cite{vaswani2017attention} have also been widely used in medical imaging because of their superior performances over CNNs.
Such spatial relationships are crucial in 3D MRI images for Alzheimer's diagnosis \cite{iaccarino2021spatial}, where understanding cross-sectional interdependencies is the key.
However, transformer-based methods have yet to see widespread use in 3D medical image diagnosis.
A primary reason is that due to a lack of inductive bias of locality, lower layers of ViT can not learn the local relations well, leading to the representation being unreliable~\cite{zhu2023understanding}.
Moreover, in 3D medical imaging, the scarcity of datasets, largely due to ethical considerations that restrict access~\cite{setio2017validation,simpson2019large}, costly annotations~\cite{yu2019uncertainty,wang2023rapid}, class imbalance challenges~\cite{yan2019holistic}, and the significant computational demands of processing high-dimensional data~\cite{tajbakhsh2020embracing}, is a notable issue. 

At the same time, these models typically treat all the dimensions in the same way. In contrast, when physicians read the MRI, they usually pay different attentions to different dimensions of the images, according to the atrophic patterns of the brain. This adaptive strategy of the physicians allows them to diagnose more efficiently and accurately.

Inspired by the above, we propose {\model}, a pure transformer-based model that leverages the captured different features from each view dimension more smartly and efficiently. 
Our goal is to classify Alzheimer’s disease (AD) and normal states in 3D MRI images. 
The overall architecture of the proposed {\model} model is shown in Fig~\ref{fig:teaser}. 
{\model} factorizes 3D MRI images into three 2D sequences of slices along axial, coronal and sagittal dimensions. 
Then we combine multiple 2D slices as input and use a 2D separate transformer encoder model to classify. At the same time, we also build attention encoders across slices from the same dimension and the attention encoders across three dimensions. These encoders can help to efficiently combine the feature information better than just keep training using the slices altogether. Benefiting from the special encoder blocks, {\model} can learn the AD pathology just using a few slices instead of inputting all 2D images, which can further reduce memory footprint.
Our contribution is as follows:

\begin{itemize}
    \item We proposed a new transformer-based architecture to solve the real-world AD diagnosis problem. 
    \item We proposed a novel cross-attention mechanism and a novel guide patch embedding, which can gather the information between slices and sequences better. 
    \item Considering the structure and difference between AD and normal MRI images, we designed the morphology augmentation methods to augment the data. 
    \item We proposed an adaptive training strategy in order to guide the attention of our model, leading the model to adaptively pay more attention to the more important dimension.
    \item Overall, we name our method {\model}, which is evaluated as the state-of-the-art performance among all the baselines while occupying minimum memory.
\end{itemize}

\section{Related Works}
\label{sec:Related}

\subsection{3D Vision Transformer}

The recent success of the transformer architecture in natural language processing~\cite{vaswani2017attention} has garnered significant attention in the computer vision domain. 
The transformer has emerged as a substitute for traditional convolution operators, owing to its capacity to capture long-range dependencies. 
Vision Transformer (ViT)~\cite{dosovitskiy2020image} introduces transformer architecture into the computer vision field and starts a craze in combining transformers and images together. Many works have demonstrated remarkable achievements across various tasks, with several cutting-edge methods incorporating transformers for enhanced learning. 

Some attention-based methods have been proposed for 3D image classification. COVID-VIT~\cite{zhou2023mdvt} uses 3D vision transformers to exploit CT chest information for the accurate classification of COVID.
I3D~\cite{carreira2017quo} proposes a new two-stream inflated 3D ConvNet to learn seamless spatio-temporal feature extractors from video, which can be used to do human action classification.
At the same time, many existing works also deal with 3D object detection problems. 
Pointformer~\cite{pan20213d} captures and aggregates local and global features together to do both indoor and outdoor object detection.
3DERT~\cite{misra2021end} proposes an encoder-decoder module that can be applied directly on the point cloud for extracting feature information, and then predicting 3D bounding boxes. 
Also, image segmentation is a hot topic in the both computer vision and medical imaging fields.
Swin UNETR~\cite{hatamizadeh2021swin} projects multi-modal input data into a 1D sequence of embedding and uses it as input to an encoder composed of a hierarchical Swin Transformer~\cite{liu2021swin}.

\textbf{Key Differences:}
These models are all using 3D architecture to deal with 3D input, which is inefficient in medical field due to the high value of medical images and limited dataset size. Unlike them, our 2D {\model} utilizes different blocks to first extract features among different slices and dimensions, then use a cross-attention mechanism to combine these features together, which can better release the abilities of transformer architecture.

\subsection{Deep Learning for Medical Image Analysis}

With the success of deep learning models, extensive research interest has been devoted to deep learning for the development of novel medical image processing algorithms, resulting in remarkably successful deep learning-based models that effectively support disease detection and diagnosis in various medical imaging tasks~\cite{chen2022recent}. 
U-Net and its variants dominate medical image analysis, which is widely used in image segmentation. Attention U-Net~\cite{oktay2018attention} incorporates attention gates into the U-Net architecture to learn important salient features and suppress irrelevant features. 

For medical image classification, AG-CNN~\cite{guan2018diagnose} uses the attention mechanism to identify discriminative regions from the global 2D image and fuse the global and local information together to better diagnose thorax disease from chest X-rays. MedicalNet~\cite{chen2019med3d} uses the resnet-based~\cite{he2016deep} model with transfer learning to solve the problem of lacking datasets. DomainKnowledge4AD~\cite{zhou2023learning} uses ResNet18 to extract high-dimensional features and proposes domain-knowledge encoding which can capture domain-invariant features and domain-specific features to help predict AD.
M3T~\cite{jang2022m3t} tries to leverage CNNs to capture the local features and use traditional transformer encoders for a long-range relationship in 3D MRI images.

\textbf{Key Differences:} These methods usually focus on CNN based model to extract and combine features, which has been outperformed by transformer-based models.
M3T tries to concate transformer blocks after CNNs, however, they propose a much bigger model and treat all slices as the same which is inefficient. In our work, we use a pure transformer-based model with different kinds of encoders to do Alzheimer's classification and have demonstrated {\model} can outperform other deep learning models in both classification accuracy results and model size.
\section{Methodology}
\label{sec:Methods}

\begin{figure*}
\centering
\includegraphics[width=0.8\textwidth]{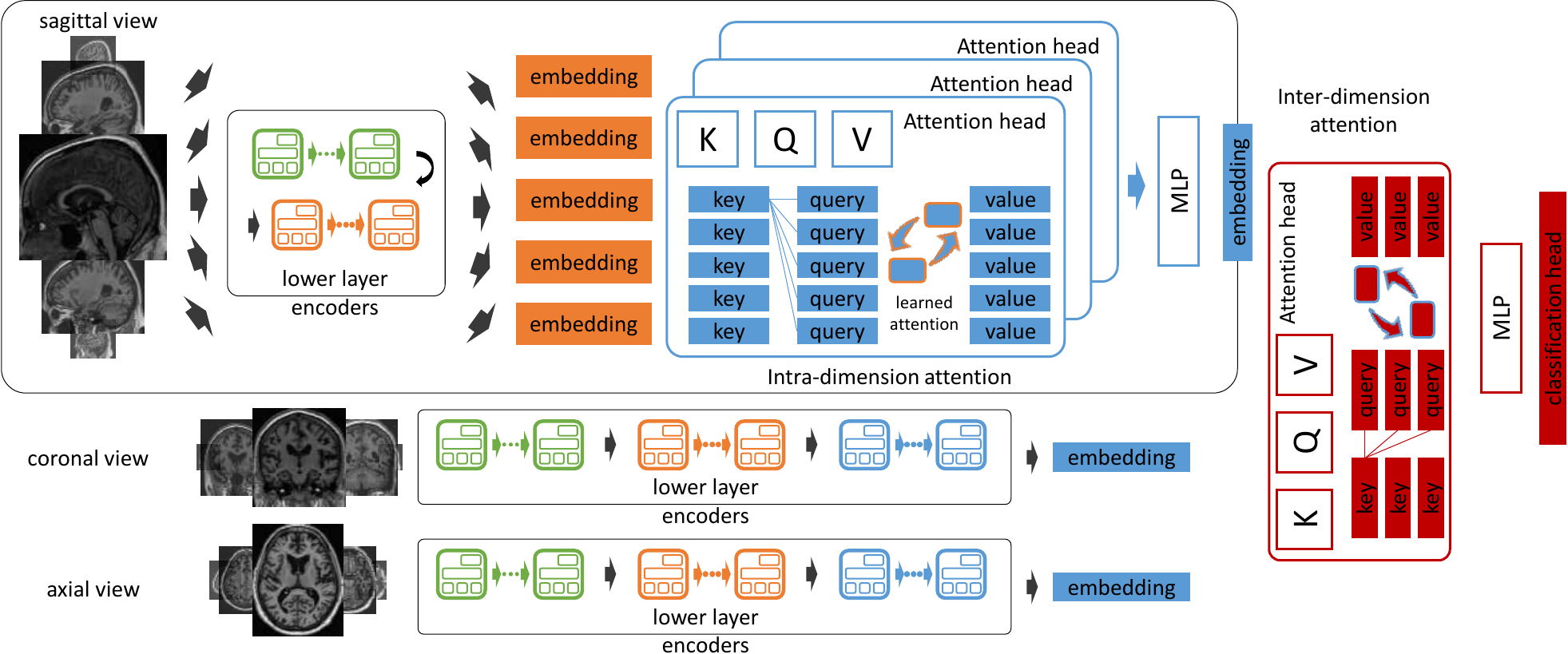}
  \caption{The detailed architecture for our {\model}. {\model} consists of four main components:
\textbf{S}elf-\textbf{A}ttention \textbf{E}ncoders (SAE) across three views,
\textbf{D}imension-specific \textbf{S}elf-\textbf{A}ttention \textbf{E}ncoders (DS-AE),
\textbf{Intra}-dimension \textbf{C}ross-\textbf{A}ttention \textbf{E}ncoders (IntraCAE),
\textbf{Inter}-dimension \textbf{C}ross-\textbf{A}ttention \textbf{E}ncoders (InterCAE). 
The figure shows more details in sagittal view for illustration purposes. In practice, the model will adaptively attend to different views.}  
  \label{fig:arch}
\end{figure*}

\subsection{ADAPT Architecture}
Our model {\model} builds upon the ViT architecture, extending it to model 3D images.
Figure~\ref{fig:arch} shows the details. 
{\model} mainly consists of 4 blocks: 
\begin{itemize}
    \item \textbf{S}elf-\textbf{A}ttention \textbf{E}ncoders (SAE) across three views
    \item \textbf{D}imension-specific \textbf{S}elf-\textbf{A}ttention \textbf{E}ncoders (DS-AE)
    \item \textbf{Intra}-dimension \textbf{C}ross-\textbf{A}ttention \textbf{E}ncoders (IntraCAE)
    \item \textbf{Inter}-dimension \textbf{C}ross-\textbf{A}ttention \textbf{E}ncoders (InterCAE)
\end{itemize}

The design inspired from a real-world setting
where physicians will pay different attentions to different views, 
according to the brain pattern. 
Thus, 
we design the encoders so that they not only can extract and fusion features from local and global but also can give different weights to different views.
To be specific, first, 
to better obtain the complete information of the 3D image, we cut each image along three views: 
sagittal view (along x-axis), coronal view (along y-axis), and axial view (along z-axis). 
We use $n$ images from each view as the model input.
Then similar to ViT, {\model} also uses the image patch and patch embedding method to embed the 2D images into 3 sequences including $3\times n$ slices with guide patch embedding layer $\textbf{x}_{guide}$, then concatenates them together as the input to the transformer encoders (Eq.~\ref{patch}).
The guide patch embedding aims to reshape the whole sequence into a sequence of flattened 2D patches that has the same shape as the sequence after the normal patch, which means the guide patch embedding has the input channel with the number $3\times n$. 
Thus, we can use 3D models to extract the global information and add it to each special slice sequence.
Because our model mainly focuses on 2D slice dimension, guide patch embedding can help to keep the relative position information of 3D brain.
\begin{equation}
\label{patch}
\begin{aligned}
    \textbf{S}_{0}=[\textbf{x}_{class}; \underbrace{\textbf{x}_{p_{1}}+\textbf{x}_{guide};\cdots;\textbf{x}_{p_{n}}+\textbf{x}_{guide}}_{sagittal}; 
\\ \underbrace{\cdots;\textbf{x}_{p_{2n}}+\textbf{x}_{guide}}_{coronal} ;\underbrace{\cdots;\textbf{x}_{p_{3n}}+\textbf{x}_{guide}}_{axial}]
\end{aligned}
\end{equation}
\begin{equation}
\label{pos}
\textbf{S}_{0}=\textbf{S}_{0}+\textbf{E}_{pos} \quad  \quad \textbf{E}_{pos} \in \mathbb{R}^{(3\cdot n\cdot N+1)\times D}
\end{equation}
Second, the lower layer encoders learn the bias attention among multiple slices and multiple views. To be more specific, the shared \textbf{S}elf-\textbf{A}ttention \textbf{E}ncoders (SAE) across three view dimensions are designed to learn not only the attention of the slice itself but also the relationship between all slices. The designed encoder can realize global information extraction for the first time. 
These encoders can also help to keep the relative position information of 3D MRI.
These networks are Siamese networks~\cite{guo2017learning} which share the same weights. 
\begin{equation}
\label{slice}
\textbf{S}_{0}^{s}=[\textbf{x}_{class}^{s};\textbf{x}_{p_{s}}]
\quad  \quad s \in (1,3\cdot n)
\end{equation}
\begin{equation}
\label{slice1}
\textbf{S}_{l}^{s}={\rm SAE( \textbf{S}_\textit{l-1}^\textit{s} )}
\quad  \quad l=1...L_{\textnormal{SAE}}
\end{equation}
The \textbf{D}imension-specific \textbf{S}elf-\textbf{A}ttention \textbf{E}ncoders (DS-AE) also aim to learn the attention of the slice itself. However, compared with SAE, these encoders focus more on the relationship between the slices from the same dimension sequence. These encoders can better extract the local features from the same view dimension. This will fill the gap that transformers cannot capture the local features well however the local embeddings of different brain tissues (such as hippocampus and cortex) are really important in AD diagnosis.
In the following equation, t means the three different views.
\begin{equation}
\label{slice1}
\begin{aligned}
\textbf{S}_{l}^{t\cdot s}={\rm DSAE_{t}} (\textbf{S}_\textit{l-1} ^{t\cdot s} {\rm)}
\quad  \quad s \in (1, n), t \in (1,3) \\
l=(L_{\textnormal{SAE}}+1)...(L_{\textnormal{SAE}}+L_{\textnormal{DSAE}})
\end{aligned}
\end{equation}
We will fusion the local features from the same dimension first.
So we design \textbf{Intra}-dimension \textbf{C}ross-\textbf{A}ttention \textbf{E}ncoders (IntraCAE). 
Here {\model} will apply cross embedding mechanism to the input embeddings. (Details are in section \ref{cross-attention}.)
After the IntraCAE, the embeddings will gather the features from different slices of the same view sufficiently.
\begin{equation}
\label{slice1}
\begin{aligned}
\textbf{S}_{l}^{t\cdot s}=&{\rm IntraCAE_{t}} (\textbf{S}_\textit{l-1} ^{t\cdot s} {\rm)}
\quad  \quad  s \in (1, n), t \in (1,3) \\
l=&(L_{\textnormal{SAE}}+L_{\textnormal{DSAE}}+1)...\\&(L_{\textnormal{SAE}}+L_{\textnormal{DSAE}}+L_{\textnormal{IntraCAE}})
\end{aligned}
\end{equation}
After combining the features between slices of the same dimension independently, the last \textbf{Inter}-dimension \textbf{C}ross-\textbf{A}ttention \textbf{E}ncoders (InterCAE) are proposed to learn the inter-dimension relationship among different sequences from different views. This is corresponding to the SAE layer and will gather the global features together. InterCAE will apply cross embedding mechanism again into the view-dependent embeddings. 
\begin{equation}
\label{slice1}
\begin{aligned}
\textbf{S}_{l}^{t}&={\rm InterCAE_{t}} (\textbf{S}_\textit{l-1}^{t} {\rm)} \quad  \quad  t \in (1,3)\\
l&=(L_{\textnormal{SAE}}+L_{\textnormal{DSAE}}+L_{\textnormal{IntraCAE}}+1)...\\&(L_{\textnormal{SAE}}+L_{\textnormal{DSAE}}+L_{\textnormal{IntraCAE}}+L_{\textnormal{InterCAE}})
\end{aligned}
\end{equation}
Finally, the $[class]$ tokens of the output from three dimensions will be averaged and sent to Layer Norm and classification MLP head as Eq.~\ref{4} and Eq.~\ref{5} to get the final diagnosis result: AD or normal.

\subsection{Fusion Attention Mechanism}
\label{cross-attention}

The above architecture will allow us to learn the intricies of AD pathologies along three different dimensions. 
However, the complicatedness of AD will 
require the model to 
thoroughly integrate the information from these three dimensions. 
Thus, we propose a cross-attention mechanism, namely funsion attention. 
The fusion attention adds the embeddings together directly. 
However, different from simply adding them together one by one, it adds the embeddings representing the patches but not the tokens.
Note that the $[class]$ token of each embedding has aggregated the information from one slice in previous encoders, so this operation will let the embeddings more focus on themselves when learning attention. 
At the same time, it can also extract the feature information from other slices or dimensions. 
The fusion attention applied to both IntraCAE and InterCAE, but here we use IntraCAE as an example:
\begin{equation}
\label{slice1}
\begin{aligned}
\textbf{S}_{l}^{t \cdot s} =\textbf{x}_{class}^{t \cdot s} \oplus (\textbf{x}_{p_{(t-1) \cdot n +1}}+
\cdots+\textbf{x}_{p_{t \cdot n}})\\
{\rm where} \quad s \in (1, n), t \in (1,3)
\end{aligned}
\end{equation}

In a more formal way,
the traditional attention mechanism is shown as Eq.~\ref{attention_tradition}. After fusing these two embeddings, the $K$ matrix of the first embedding will consist of the $K$ value corresponding to the $[class]$ token from the first embedding, and the $K$ matrix corresponding to fusion embedding, 
similarly for $Q$ matrix.
After the matrix calculation, Eq.~\ref{fuse}
fuses the information from two embeddings while keeping some unique information from the special $[class]$ token.
\begin{equation}
\label{attention_tradition}
\begin{aligned}
H=softmax(\frac{QK^T}{\sqrt{d_k}})V
\end{aligned}
\end{equation}
\begin{equation}
\label{K_total}
\begin{aligned}
K_1=[K_{class_{1}},K_1+K_2],
Q_1=[Q_{class_{1}},Q_1+Q_2]
\end{aligned}
\end{equation}
\begin{equation}
\label{fuse}
Q_{1}K_{1}^{T}=\begin{bmatrix} Q_{class_{1}}K_{class_{1}} & (Q_1+Q_2)K_{class_{1}} \\ Q_{class_{1}}(K_1+K_2) & (Q_1+Q_2)(K_1+K_2) \end{bmatrix}
\end{equation}
\subsection{Morphology Augmentation}
A key characteristic of the AD-plagued brain is that, 
as the disease progresses, an increasing amount of brain mass will suffer from atropy.
When this process is reflected in brain imaging, 
the there will be empty ``holes'' of the brain if one has AD.
Based on this, we propose a morphology augmentation, 
an augmentation method which help to expand and reduce the size of the atrophy, causing the improvement of the model.
This augmentation is based on atrophy expansion and atrophy reduction shown in Eq. ~\ref{erosion},~\ref{dilation}. $f$ is the input image, $b_{N}$ is the atrophy expansion or atrophy reduction element, $(x,y)$ and $(s,t)$ are the coordinates in $f$ and $b_{N}$ respectively.
\begin{equation}
\label{erosion}
\begin{aligned}
\relax[f\ominus b_{N}](x,y)=\min\limits_{(s,t)\in b_N} \left\{ f(x+s,y+t)-b_N(s,t)  \right\}
\end{aligned}
\end{equation}
\begin{equation}
\label{dilation}
\begin{aligned}
\relax[f\oplus b_{N}](x,y)=\max\limits_{(s,t)\in b_N} \left\{ f(x-s,y-t)+b_N(s,t)  \right\}
\end{aligned}
\end{equation}

\begin{figure}
\centering
\includegraphics[scale=0.3]{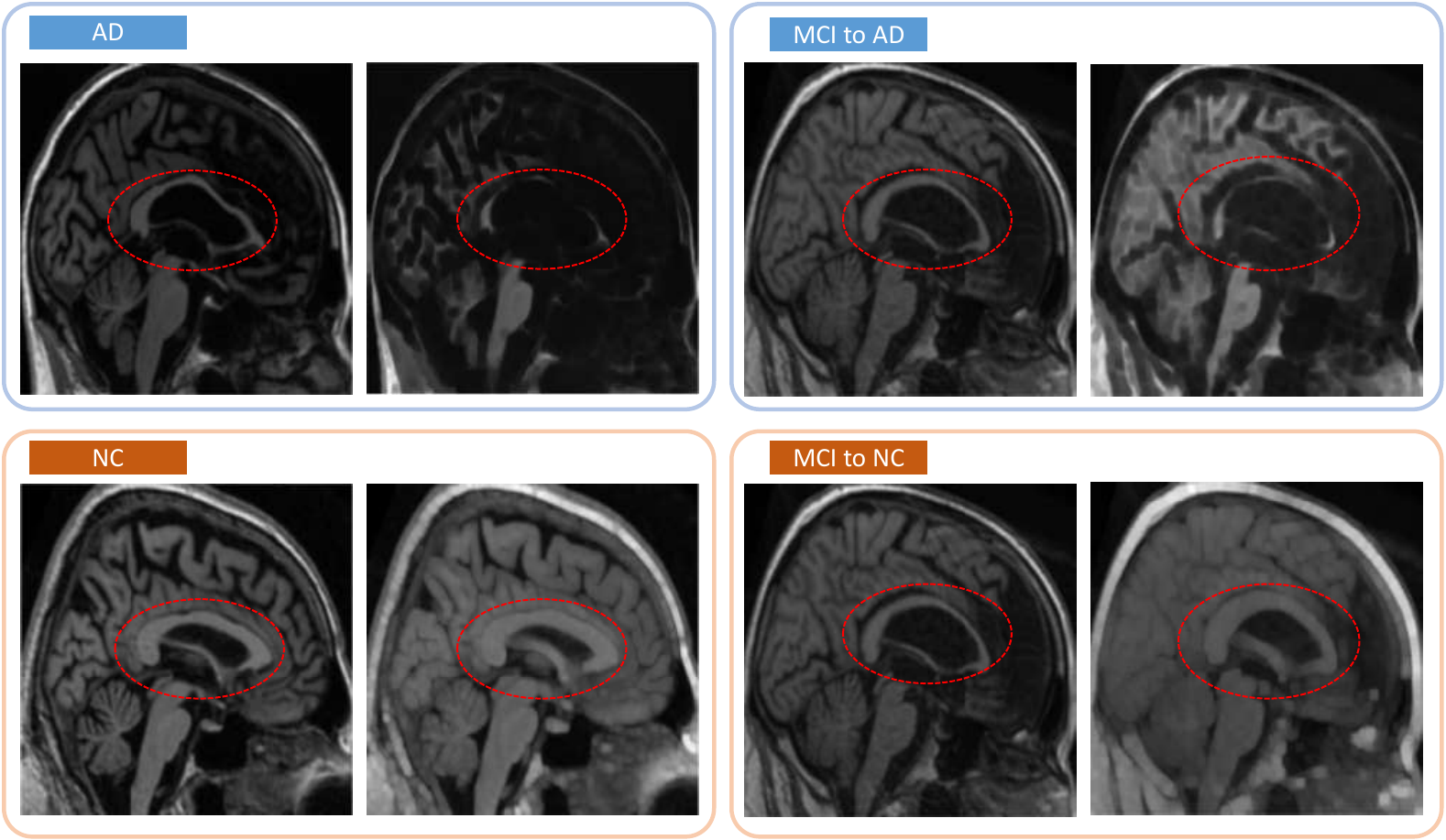}
  \caption{The visualization of Alzheimer's Disease (AD) image, Normal Control (NC) image and  Mild Cognitive Impairment (MCI) image. The left is the raw image and the right is the augmented image. The cerebral ventricle (red circle) has a significant difference in size for AD and NC. }
  \label{fig:aug}
\end{figure}

We apply atropy expansion augmentation to AD images and MCI images and label the resultant images as AD; 
on the other hand, we apply atropy reduction augmentation to Normal Control(NC) images and MCI images and label the resultant images as NC, where MCI is the prodromal stage of AD.
The visualization of morphology augmentation is shown in Fig.~\ref{fig:aug}.

\subsection{Adaptive Training Strategy}
To further investigate the potential of {\model}, we propose an attention score based training strategy in order to allow our model to extract more features from the more important dimension with limited size of inputs. 
We calculate the attention score of each dimension after the final inter-dimension cross attention encoder layer according to Eq.~\ref{attention_special}. Because our $[class]$ token is dimension specific, so we just calculate the attention score of the $[class]$ token as the representation of the special dimension. This strategy allows our network to adaptively choose the slice number of each dimension while updating itself.
\begin{equation}
\label{attention_special}
\begin{aligned}
H_{dim}=softmax(\frac{Q_{class_{dim}}K^T}{\sqrt{d_k}})V
\end{aligned}
\end{equation}
\begin{algorithm}[ht]
\renewcommand{\algorithmicrequire}{\textbf{Input:}}
\renewcommand{\algorithmicensure}{\textbf{Output:}}
\caption{{\model} Training Strategy}
\label{alg:select}
\begin{algorithmic}[1] 
\REQUIRE {3D MRI Training set $T$, initial slice number list $\psi$, model {\model} $\Theta$, total slice number $n_{total}$}
\ENSURE {Updated model $\Theta$, final list $\psi$}
\WHILE {$Training$}
\STATE Sample 2D data $\delta$ with $T$ and $\psi$ 
\STATE \textsc{Train}($\Theta$,$\delta$)
\STATE Calculate score using Eq.~\ref{attention_special} for three dimensions
\STATE Calculate cross-entropy loss and update $\Theta$
\IF{p}
\STATE Update $\psi$ according to Eq~\ref{range},~\ref{range0}.
\ELSE
\STATE \textsc{Initialize}(\textit{$\psi$})
\ENDIF
\ENDWHILE
\end{algorithmic}
\end{algorithm}

We then adaptively update the slice number of each dimension based on normalized attention scores using 
  Eq.~\ref{range0},~\ref{range}, where $n$ is the total slice number and $\psi$ is the slice number list. Here we also constrain the selection pool to make sure the model will attend across multiple attentions. The full training strategy is shown in algorithm~\ref{alg:select}. 
To avoid the model will stick with certain view dimension after the first choice, we also allow the model to change the attention with certain probabilities p. 
\begin{equation}
\label{range0}
\begin{aligned}
n_{dim}=round(\frac{\hat{H_{dim}}}{\sum_{r \in \textsc{$\psi$}} r}*n_{total})
\end{aligned}
\end{equation}
\begin{equation}
\label{range}
    n_{dim}=
        \begin{cases}
            n_{min},n_{dim}\geq n_{min}\\
            n_{max},n_{dim}\leq n_{max}
        \end{cases}
\end{equation}
\section{Experiments}
\label{sec:Experiments}

\begin{table*}\centering
\begin{tabular}{@{}lrrccccc@{}}
\toprule
\multicolumn{1}{c}{\multirow{2}{*}{Model name}} &
  \multicolumn{1}{c}{\multirow{2}{*}{\begin{tabular}[c]{@{}c@{}}Model size\\  (\#params)\end{tabular}}} &
  \multicolumn{1}{l}{\multirow{2}{*}{GFLOPs}} &
  \multicolumn{2}{c}{ADNI} &
  AIBL &
  MIRIAD &
  OASIS \\ \cmidrule(l){4-8} 
\multicolumn{1}{c}{} &
  \multicolumn{1}{c}{} &
  \multicolumn{1}{l}{} &
  \multicolumn{1}{l}{val acc.} &
  \multicolumn{1}{l}{test acc.} &
  \multicolumn{1}{l}{test acc.} &
  \multicolumn{1}{l}{test acc.} &
  \multicolumn{1}{l}{test acc.} \\ \midrule
MedicalNet-10~\cite{chen2019med3d}   & 17,723,458  & 225.7  & 0.827 & 0.843 & 0.856 & 0.847 & 0.793 \\
MedicalNet-18~\cite{chen2019med3d}   & 36,527,938  & 492.6  & 0.756 & 0.793 & 0.820 & 0.800 & 0.809 \\
MedicalNet-34~\cite{chen2019med3d}   & 66,837,570  & 910.8  & 0.571 & 0.669 & 0.847 & 0.782 & 0.585 \\
MedicalNet-50~\cite{chen2019med3d}   & 59,626,818  & 666.8  & 0.588 & 0.471 & 0.303 & 0.706 & 0.317 \\
MedicalNet-101~\cite{chen2019med3d}  & 98,672,962  & 1181.1 & 0.558 & 0.425 & 0.320 & 0.656 & 0.291 \\
MedicalNet-152~\cite{chen2019med3d}  & 130,831,682 & 1604.8 & 0.564 & 0.381 & 0.175 & 0.662 & 0.239 \\
3D ResNet-34~\cite{he2016deep}    & 63,470,658  & 341.1  &0.506  & 0.618 &   0.779 & 0.453 & 0.745 \\
3D ResNet-50~\cite{he2016deep}    & 46,159,170  & 256.9  & 0.519 & 0.618 &   0.776 & 0.657 & 0.720 \\
3D ResNet-101~\cite{he2016deep}   & 85,205,314  & 391.1  & 0.569 & 0.526 &   0.601 & 0.426 & 0.433 \\
3D DenseNet-121~\cite{huang2017densely} & 11,244,674  & 260.5  & 0.609 & 0.675 & 0.816 & 0.408 & 0.772 \\
3D DenseNet-201~\cite{huang2017densely} & 25,334,658  & 286.8  & 0.564 & 0.612 & 0.672 & 0.553 & 0.671 \\
Knowledge4D~\cite{zhou2023learning}     & 33,162,880  & 633.9  & 0.603 & 0.728 & 0.845 & 0.658 & 0.784 \\
I3d~\cite{carreira2017quo}             & 12,247,332  & 191.0  & 0.513 & 0.629 & 0.674 & 0.448 & 0.607 \\
FCNlinksCNN~\cite{qiu2020development}     & 310,488,372 & 375.6  & 0.423 & 0.614 & 0.708 & 0.370 & 0.691 \\
COVID-ViT~\cite{gao2021covid}       & 78,177,282  & 448.6  & 0.506 & 0.573 & 0.644 & 0.662 & 0.715 \\
Uni4Eye~\cite{cai2022uni4eye}         & 340,324,866 & 78.4   & 0.622 & 0.646 & 0.697 & 0.647 & 0.688 \\
\textbf{ADAPT} &
  \textbf{9,695,490} &
  \textbf{46.3} &
  \textbf{0.904} &
  \textbf{0.924} &
  \textbf{0.910} &
  \textbf{0.903} &
  \textbf{0.817} \\ \bottomrule
\end{tabular}
\caption{Comparison of accuracy various 3D CNN-based and transformer-based models on multi-institutional Alzheimer’s disease dataset.}
\label{tab_acc}
\end{table*}

\subsection{Experimental Dataset}
To verify the effectiveness of our {\model}, we use the dataset from the Alzheimer’s Disease Neuroimaging Initiative (ADNI) for the training process. 
This dataset consists of MRI images of T1-weighted magnetic resonance imaging subjects.
There are a total of 3,891 3D MRI images in the dataset, including 1,216 normal cases (NC), 1,110 AD cases and 1,565 MCI cases.
During the training, 878 normal images, 884 AD images and 1565 MCI images were split into the training set, with 72 normal images and 81 AD images as a validation set, together with 266 normal images and 145 AD images as a testing set.

At the same time, to evaluate the performance of our {\model} and other deep learning baseline models, we also consider other datasets as test sets. We mainly acquire them from three other institutions with the ADNI test dataset: Australian Imaging, Biomarker and Lifestyle Flagship Study of Ageing (AIBL), Minimal Interval Resonance Imaging in Alzheimer's Disease (MIRIAD), and The Open Access Series of Imaging Studies (OASIS). 
The AIBL dataset contains a total of 413 images with 363 NC and 50 AD after dropping all MCI cases.
The MIRIAD dataset contains a total of 523 cases which consist of 177 NC and 346 AD cases. 
The OASIS dataset contains a total of 2157 cases which consist of 1692 NC and 465 AD cases. 
Each of the images for any dataset is a 3D grayscale image.

\subsection{Implementation Details}
We implement consistent data pre-processing techniques to normalize and standardize MRI images sourced from a multi-institutional database.
We first do data augmentation in the following steps. we have followed closely the recommended protocol from the medical community \cite{wen2020convolutional} 
to process the data.  
Firstly, we do bias field correction with N4ITK method~\cite{tustison2010n4itk}. Next, we register each image to the MNI space~\cite{fonov2009unbiased,fonov2011unbiased} with the ICBM 2009c nonlinear symmetric template by performing a affine registration using the SyN algorithm~\cite{avants2014insight} from ANTs~\cite{avants2008symmetric}. At the same time, the registered images were further cropped to remove the background to improve the computational efficiency. These operations result in 1 mm isotropic voxels for each image. Intensity rescaling, which was performed based on the minimum and
maximum values, denoted as MinMax, was also set to be optional to study its influence on the classification results.
Finally, the deep QC system~\cite{fonov2018deep} is performed to check the quality of the linearly registered data. The software outputs a probability indicating how accurate the registration is. We excluded the scans with a probability lower than 0.5. Overall, the registration process we perform on the data maps different sets of images into a single coordinate system to prepare the data for our later usage.

We also use the Torchio library~\cite{perez2021torchio} in the training set.
At the same time, we resize all the MRI images with Scipy library~\cite{virtanen2020scipy} into 224$\times$224$\times$224 to better fit the input of our {\model}. 
Finally, we employed the zero-mean unit-variance method to normalize the intensity of all voxels within the images.

For the training dataset, 
we apply morphology augmentation to the same MCI data, classify the MCI into NC after doing atrophy reduction augmentation, and classify it into AD after doing atrophy expansion augmentation. In this way, each MCI is used twice, significantly enlarging the dataset. At the same time, we also do morphology augmentation to AD and NC images randomly, with a probability of 0.5.

After preprocessing the 3D MRI images, we cut them into 2D slices along sagittal, coronal and axial views. Then we choose 16 slices in each view as the initial data and concatenate them into a sequence. 
We choose equidistant slices on each view and embed them into patch embedding similar to ViT. 
Here we choose the embed layer from ~\cite{touvron2022three}. 
Then we use a total of 6 standard transformer attention layers, and 1 layer for each of the first two encoders, 2 layers for each of the last two encoders, with 4 heads. 
For the adaptive training strategy, we set the probability p as 0.8.
At last, because we have three $[class]$ tokens, each representing a special view dimension, we use a classification MLP head, with input feature number 3$\times$256 and output feature number 2, aiming to figure out whether the image is from a disease or not.

We implemented {\model} using a Pytorch library~\cite{paszke2019pytorch}. {\model} was trained using an AdamW optimizer with a learning rate of 0.00005. All other parameters are default. At the same time, we also took the advantage of cosine learning rate from~\cite{loshchilov2016sgdr}.  
We treat this as a binary classification task, so we use cross-entropy loss~\cite{zhang2018generalized}.
The training process used 2 80G NVIDIA A800 GPUs. Due to the memory capacity, we use 6 batches on each GPU, meaning a total batch size of 12.

\subsection{Evaluation Between Baselines}
Our {\model} was compared with various baseline models, including 3D CNN-based models: 3D DenseNet (121, 201)~\cite{huang2017densely}, 3D ResNet (34, 50, 101)~\cite{he2016deep} because they have been widely used for
AD classification~\cite{korolev2017residual,ruiz20203d,yang2018visual,zhang20213d}. We also add other baselines to show the capability of our {\model}, including: MedicalNet~\cite{chen2019med3d}, I3D~\cite{carreira2017quo}, FCNlinksCNN~\cite{qiu2020development} and Knowledge4D~\cite{zhou2023learning}.
There are various versions of MedicalNet, each of which is based on a basic Resnet~\cite{he2016deep} model, such that MedicalNet-10 is based on Resnet-10 respectively.
We also compare our method with 3D transformer-based models: COVID-VIT~\cite{gao2021covid}, Uni4Eye~\cite{cai2022uni4eye}. 

In the experiment, we chose a total of 48 slices as input, meaning 16 equidistant slices on each view as initial. 
Because we found that the central part of the 3D images would be more important and consist of more useful information, we applied the \textbf{important sampling} method in our slice-picking stage. To be more specific, for a 224$\times$224$\times2$24 image, we pick equidistant slices from $\rm 52^{nd}$ to $\rm 172^{nd}$ on each view.

The quantitative performance is presented in Table \ref{tab_acc}. We chose the model with the best validation accuracy on ADNI and then tested it on various Alzheimer's disease datasets.
We also recorded the total parameters and GFLOPs of each model.
Overall, {\model} achieves the best performance on i.i.d testing scenario (ADNI) 
as well as all out-of-domain testing scenarios (AIBL, MIRIAD and OASIS). 
We believe these results show that {\model} is not only superior 
in Alzheimer's diagnosis in i.i.d setting, 
but also fairly robust when the testing data is collected from different facilities. 
At the same time, our model has the least parameters and GLOPs, demonstrating the success of our novel method in attacking the AD diagnosing task using 2D based model.

The best performance is achieved when {\model} chooses 10 slices from saggital view, and 19 slices from the coronal and axial view respectively. As compared with table~\ref{tab_models}, we found the interesting facts that coronal and axial view may contain more differential relationships about cortex and ventricle of AD and NC, which can help the model learn the special attention features accurately.

\subsection{Ablation Study}

\begin{table}[]
\scalebox{0.75}{
\begin{tabular}{cccccc}
\hline
\multirow{2}{*}{Layer Number} & \multicolumn{2}{c}{ADNI} & AIBL & MIRIAD & OASIS \\ \cline{2-6} 
 & Val acc. & Test acc. & Test acc. & Test acc. & Test acc. \\ \hline
1+1+1+1 & 0.713 & 0.776 & 0.800 & 0.685 & 0.793 \\
2+2+1+1 & 0.770 & 0.811 & 0.863 & 0.903 & 0.716 \\
2+2+2+2 & 0.881 & 0.911 & 0.897 & 0.669 & 0.800 \\
3+3+3+3 & 0.917 & 0.895 & 0.907 & 0.723 & 0.806 \\
\textbf{Ours (1+1+2+2)} & 0.904 & \textbf{0.924} & \textbf{0.910} & \textbf{0.903} & \textbf{0.817} \\ \hline
\end{tabular}}
\caption{Comparison of accuracy between {\model} and three variants ablating with different numbers of transformer layers in each encoder in the four datasets.}
\label{tab_layer}
\end{table}

\begin{table}[]
\scalebox{0.67}{
\begin{tabular}{@{}cccccc@{}}
\toprule
\multirow{2}{*}{Cross-Attention Mechanism} & \multicolumn{2}{c}{ADNI} & AIBL & MIRIAD & OASIS \\ \cmidrule(l){2-6} 
 & Val acc. & Test acc. & Test acc. & Test acc. & Test acc. \\ \midrule
No Cross-Attention & 0.719 & 0.627 & 0.810 & 0.710 & 0.675 \\
Class Token Cross-Attention & 0.878 & 0.848 & 0.864 & 0.709 & 0.606 \\
Easy Concat Cross-Attention & 0.917 & 0.783 & 0.723 & 0.806 & 0.681 \\
\textbf{Ours (Fusion Attention)} & 0.904 & \textbf{0.924} & \textbf{0.910} & \textbf{0.903} & \textbf{0.817} \\ \bottomrule
\end{tabular}}
\caption{Comparison of accuracy between {\model} and three variants ablating different cross attention mechanisms in the four datasets.}
\label{tab_cross}
\end{table}

\begin{table}[]
\scalebox{0.65}{
\begin{tabular}{@{}cccccc@{}}
\toprule
\multirow{2}{*}{Models} & \multicolumn{2}{c}{ADNI} & AIBL & MIRIAD & OASIS \\ \cmidrule(l){2-6} 
 & Val acc. & Test acc. & Test acc. & Test acc. & Test acc. \\ \midrule
w/o Morphology Augmentation & 0.604 & 0.738 & 0.889 & 0.828 & 0.792 \\
w/o Adaptive Training & 0.788 & 0.855 & 0.883 & 0.902 & 0.807 \\
w/o Guide Eembedding & 0.750 & 0.800 & 0.863 & 0.869 & 0.813 \\
w/o Torchio & 0.678 & 0.755 & 0.825 & 0.664 & 0.783 \\
w/o Pretrained Weights & 0.731 & 0.738 & 0.877 & 0.746 & 0.800 \\
w/o Important Sampling & 0.713 & 0.712 & 0.757 & 0.767 & 0.787 \\
\textbf{ADAPT} & 0.904 & \textbf{0.924} & \textbf{0.910} & \textbf{0.903} & \textbf{0.817} \\ \bottomrule
\end{tabular}}
\caption{Comparison of accuracy between {\model} and five variants ablating different training augmentation settings in the four datasets.}
\label{tab_models}
\end{table}

To evaluate how effective each block is, we compared our {\model} with other variants, changing one setting each time. 
We first changed the transformer attention layers of each encoder. 
We investigate how the number of layers will affect our {\model} performance. The results are shown in Table~\ref{tab_layer}, there are four numbers in each variant, each one corresponding to an encoder block. Such as 1+1+2+2 meaning that the shared self-attention encoders, dimension-specific self-attention encoders, intra-dimension cross-attention encoders and inter-dimension cross-attention encoders have 1, 1, 2, 2 transformer attention layer respectively. The result shows that {\model} outperforms all the variants on test accuracy in three out of four datasets (ADNI, AIBL, MIRIAD) while using small enough memory.

Table~\ref{tab_cross} shows how different cross-attention mechanisms will affect the final result. The first variant: No Cross-Attention, meaning that we didn't apply any cross-attention mechanism in the last two encoder blocks. Class Token Cross-Attention is a variant of Eq.~\ref{K_total}. It adds the $[class]$ token embedding up but not the embedding behind the $[class]$ token. For the easy concat cross-attention mechanism, it simply concatenates the embeddings from different slices and view dimensions into a whole large embedding. Our proposed Fusion Attention achieves more than 7\% improvements to the ADNI test result while demonstrating superiority on other testing datasets, verifying that fusion attention cannot only fuse the information while keeping the unique information in each embedding.

Table~\ref{tab_models} shows other variables in our settings. We delete one important setting in each variant to see the results.
{\model} outperforms all variant models in all four datasets by 4.9\%, 2.1\%, 0.1\% and 1.0\%, respectively. The results show the great capability of different settings in augmenting the model learning ability to classify 3D MRI. 

\subsection{Visualization Result}
\begin{figure}
\centering
\includegraphics[width=0.45\textwidth]{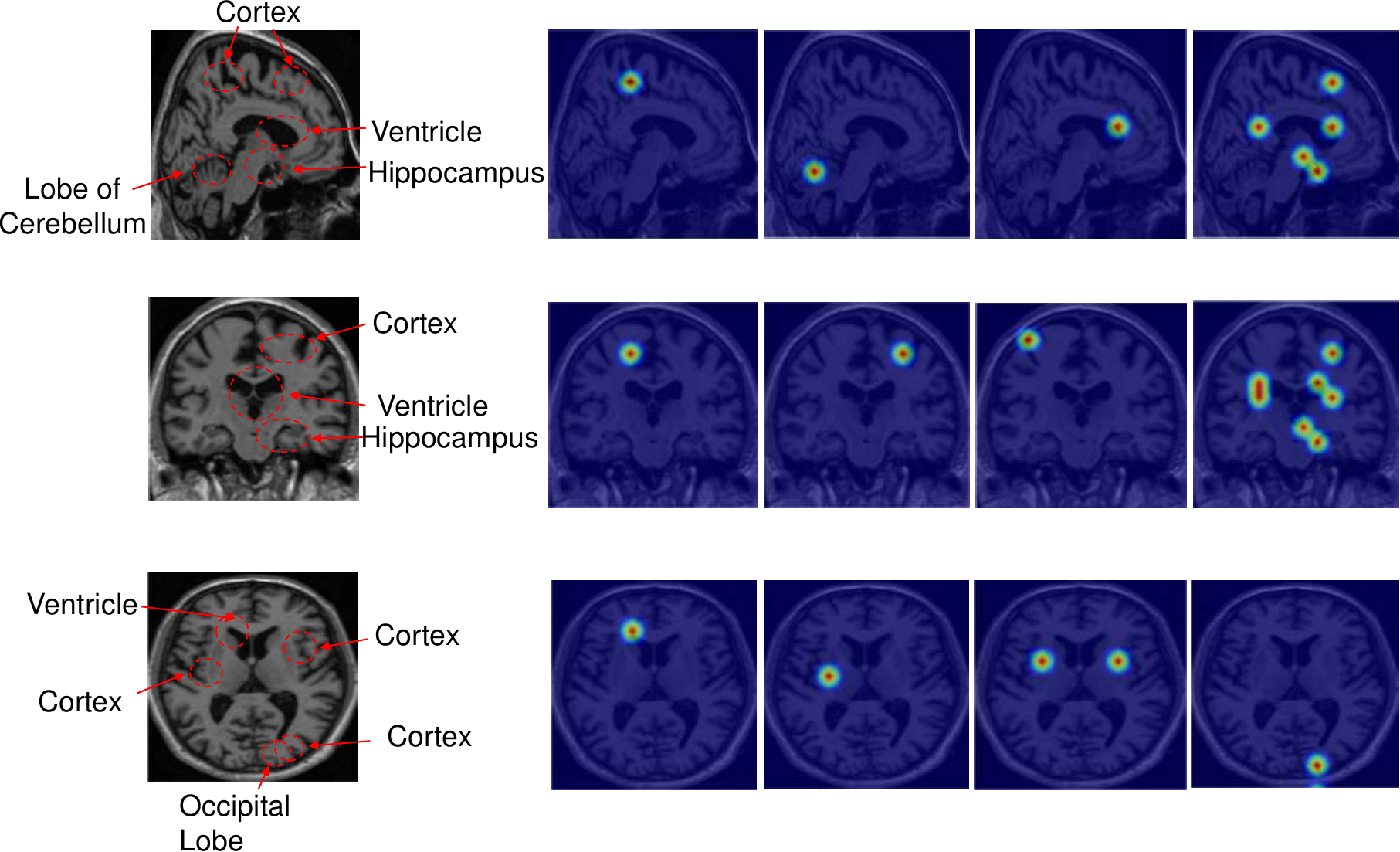}
  \caption{Attention map for Normal Control result. Each line corresponds to one view dimension: saggital, coronal and axial. }
  \label{fig:attention_map1}
\end{figure}

We visualize the activated area our model focusing on based on the transformer attention map. Figure~\ref{fig:attention_map1} shows a NC-related attention map in 3D MRI images from ADNI dataset in sagittal, coronal and axial views. Because {\model} has 4 special encoders, we visualize the attention result after each encoder.

\begin{figure}
\centering
\includegraphics[width=0.47\textwidth]{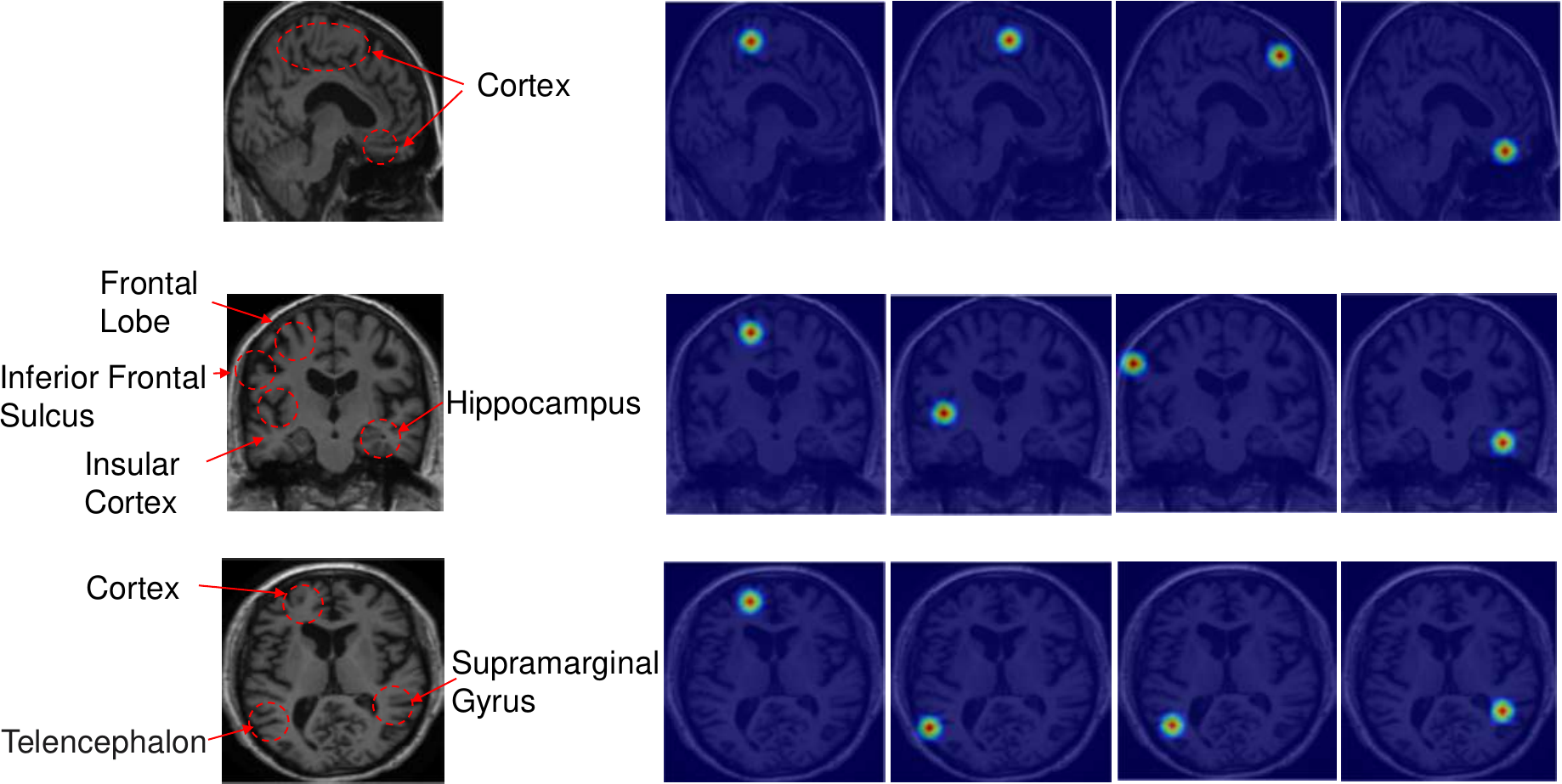}
  \caption{Attention map for Alzheimer's Disease result. Each line corresponds to one view dimension: saggital, coronal and axial.}
  \label{fig:attention_map2}
\end{figure}

We found that for NC and AD result, the attention mostly focused on some special brain tissues, such as hippocampus, cortex, ventricle and frontal lobe. 
Disruption of the frontal lobes and its associated networks are a common consequence of neurodegenerative disorders~\cite{sawyer2017diagnosing}, as well as the hippocampus is most notably damaged by AD~\cite{xu2021remodeling}.
Based on these understandings of Alzheimer's pathology~\cite{frisoni2010clinical}, {\model} successfully captured the AD-related part because with the procedure of Alzheimer's, the hippocampus and cortex begin to atrophy, and the ventricle begins to expand, which can serve as an evidence of morphology augmentation and confirm the reliability of our proposed {\model}. 
\section{Conclusions}
\label{sec:Conclusions}

We proposed a 3D medical image classification model, called {\model}, that uses various 2D transformer encoder blocks for Alzheimer’s disease diagnosis.
The proposed method uses shared self-attention encoders across different view dimensions, dimension-specific self-attention encoders, intra-dimension cross-attention encoders, and inter-dimension cross-attention encoders to extract and combine information from high-dimensional 3D MRI images, with novel techniques such as fusion attention mechanism and morphology augmentation.
With different encoders, our adaptive training strategy can allow physicians to pay more attention to different dimensions of MRI images.
The experiments show that {\model} can achieve outstanding performance while utilizing the least memory compared to various 3D image classification networks in multi-institutional test datasets. 
The visualization results show that {\model} can successfully focus on AD-related regions of 3D MRI images, guiding accurate and efficient clinical research on Alzheimer's Disease.

\section*{Impact Statements}
This paper presents work whose goal is to advance the field of Machine Learning and Alzheimer's Diagnosis. We hope our work can provide insights for medical doctors and help them diagnose AD in a convenient way. We suggest that our model should be used under human supervision to ensure a perfect result. There are many other potential societal consequences of our work, none which we feel must be specifically highlighted here.

\nocite{langley00}


\newpage
\appendix
\onecolumn



\end{document}